# Rain from Solar Scattering

Aya Thompson, ORCID: 0000-0002-1046-2052

**Abstract**

Herein we propose a method to mimic natural processes for the creation of precipitation, in a safe, economically feasible manner anywhere in the world. We propose this is accomplishable via changing the target of the well established field of aerosol dispersal for large scale climate cooling from long term cooling to short term, locally targeted dispersal. We show that such methods could induce precipitation anywhere with sufficient humidity and other conditions, and could be accomplishable at low cost with low or no safety concerns.

**Significance:** Drought and heatwaves threaten human and natural habitation around the world. We show that by copying the natural phenomena of white dust in the lower atmosphere, we suggest the harmless and globally found mineral calcium carbonate, or on oceanic coasts sea salt, we should be able to significantly cool any given region cheaply, quickly, and on demand, including to the point of inducing rain over a given area.

**Introduction**

Increasing global temperatures due to climate change have a high cost in both humanitarian(1) and economic(2) terms; including areas of increased water shortage. Multiple mitigation strategies have been proposed to counter such costs. Herein we will concentrate on Solar Radiation Management, or reduction of solar radiation heating in the earth's atmosphere, to affect artificial precipitation over a target area in order to alleviate water shortages.

**Prior Work**

Often named Stratospheric Aerosol Injection, aerosol dispersal involves the dispersal of reflective particles into the upper atmosphere, often involving sulfates, calcium carbonate, or other salts. The resulting particles mimic the natural occurrence of such after large volcanic eruptions and other phenomena, and scatter solar radiation back into space, increasing the albedo (reflectivity) of earth and thus reducing solar radiation heating(3).

Objections to deployment of such a strategy are numerous. A careful review of available literature found many are either manageable or have no basis in reality(4). However our proposal avoids most of these objections anyway by aiming for the aerosol to only be present at lower altitudes for a very short time.

Deployment strategies for aerosols are numerous. From modified cannons and anti-aircraft guns, to balloons, and more there's a wide variety of proposals. A recent review found jet airplane deployment from existing airliner type bodies to be the most cost efficient, at a cost of around $1000 USD per ton(5).

As for previous attempts at inducing artificial precipitation, the most common and well studied is that of cloud seeding(6). Here “nucleation” particles are spread into the atmosphere, around which water droplets form and eventually grow into rain. However despite being tested in numerous locations around the world for decades, effectiveness has been found to be limited, when it rises above the error bar at all. As we’ll see, aerosol cloud interaction is far more complex than the cloud seeding strategy assumes.

A close analogue to our proposal is marine cloud brightening, by which stratocumulus clouds are seeded with salts in order to increase their albedo(7). This has a long history of study, providing valuable data for our proposal. though it should be noted the goals include avoiding precipitation, and thus there are limits to how well this previous work applies.

**Overview**

We propose that precipitation can be induced over an area from a rapid cooling effect using aerosols. We’ll be concentrating on two classes of aerosols, one made solely of calcium carbonate, and the other termed “sea salts” which will be left for further study. Other possibilities will be mentioned as well for future research.

Calcium carbonate (CaCO3) is self descriptive it, is one of the most common compounds on the planet. Consequently it demonstrates no known toxicity to any living organism on the planet at relevant quantities, and nanoparticle tests on standard test organisms support this(8). However as we’ll discuss below, while CaCO3 is highly relevant for cooling applications and thus prevention of heatwaves, how useful it is for artificial precipitation may be dependent on testing.

Our other compound, one more promising for precipitation formation is sea salt. “Sea salt” is a common term covering salts naturally found in a marine environment. While the salt components anions and cations are bound via hydrogen bonding in the ocean they dissociate easily and recrystallize to salts under common circumstances. Sea salts have been studied in detail for the purposes of aerosolization, cooling, and their effects on precipitation(9,10). There are safety concerns over some of these compounds, but these have been studied in detail, such as long term deposition from natural aerosolization, which is mostly found in coastal regions(11,12). While drift of salt aerosols is well studied concerns over health and environmental impact may limit usefulness until after practical demonstrations.

We’ll start with how precipitation forms naturally from temperature differences, atmospheric water saturation, and aerosols. Detailed discussions of human and environmental safety will follow, and then a specific overview of cooling, precipitation formation, and aerosol properties of our compounds. Finally in results we’ll be estimating how these compounds can be used to intentionally create precipitation via simulations and brief estimates of costs and beneficial results.

**Methods**

Background on Precipitation Formation:

We start with cloud and precipitation formation. Earth's atmosphere is saturated with water, found in solution. In a general model clouds start to form when the temperature of a given body of air falls below its "dew point". This is defined as the temperature below which air can no longer hold more water; it is dependent on current water saturation(humidity) and temperature, as warm air can hold more water than colder air. Once below the dew point temperature water will start to form on any available "nucleation" site(6). The nucleation sites are termed "cloud condensation nucleation" (CCN), for which differing chemical makeup also affects the point at which water comes out of solution. A general overview of all this can be found here(13). Next we'll be going into detail.

CCN has a large effect on precipitation formation directly, on localized thermodynamics, and can be a primary driver or limiter of precipitation. In specifics water will come out of solution from the atmosphere not when absolutely no more water can be in the atmosphere (temperature drops below dew point) but when temperature drops below the "relative humidity of deliquescence" (DRH) of available nucleation sites(14); which is referred to as Köhler Theory after its discovered. Once there those droplets, now cloud droplets, will increase in size to the limit of their "hygroscopic growth factor" (HGF), defined as the ratio of droplet size to solute size, as long as the relative humidity keeps rising(humidity/temperature). Or once the drop is heavy enough, either from its own growth or more likely from colliding and combining with other droplets, it falls to the ground as precipitation. This is a basic overview from the starting point of CCN, below we'll go into detail.

Once cloud formation begins we see a number of other effects. First we see droplet nucleation releasing latent heat due to it being an exothermic reaction. This increase in local temperature and lowered solution water content then decreases both relative and absolute humidity. This can halt the growth of clouds and eventual precipitation, especially if CCN are numerous, generating many tiny droplets too small to collide with each other that no longer grow. Exceptions exist however. If humidity is high enough to continue growth then precipitation can still form. Thus while (15) reports an increase in precipitation from high CCN, multiple other studies found a decrease in precipitation due to smaller, more numerous droplets (13). For precipitation to form, there needs to be the correct ratio between CCN DRH, humidity, and temperature. As controlling for water saturation is difficult, we choose to concentrate on controlling CCN and temperature.

To back up, we mentioned latent heat release once water comes out of solution. This can cause an increase in atmospheric convection, raising the remaining water saturated air up to an altitude where temperature may dip below the DRH again, freezing out this water as well and further increasing cloud coverage and the possibility of precipitation(15).

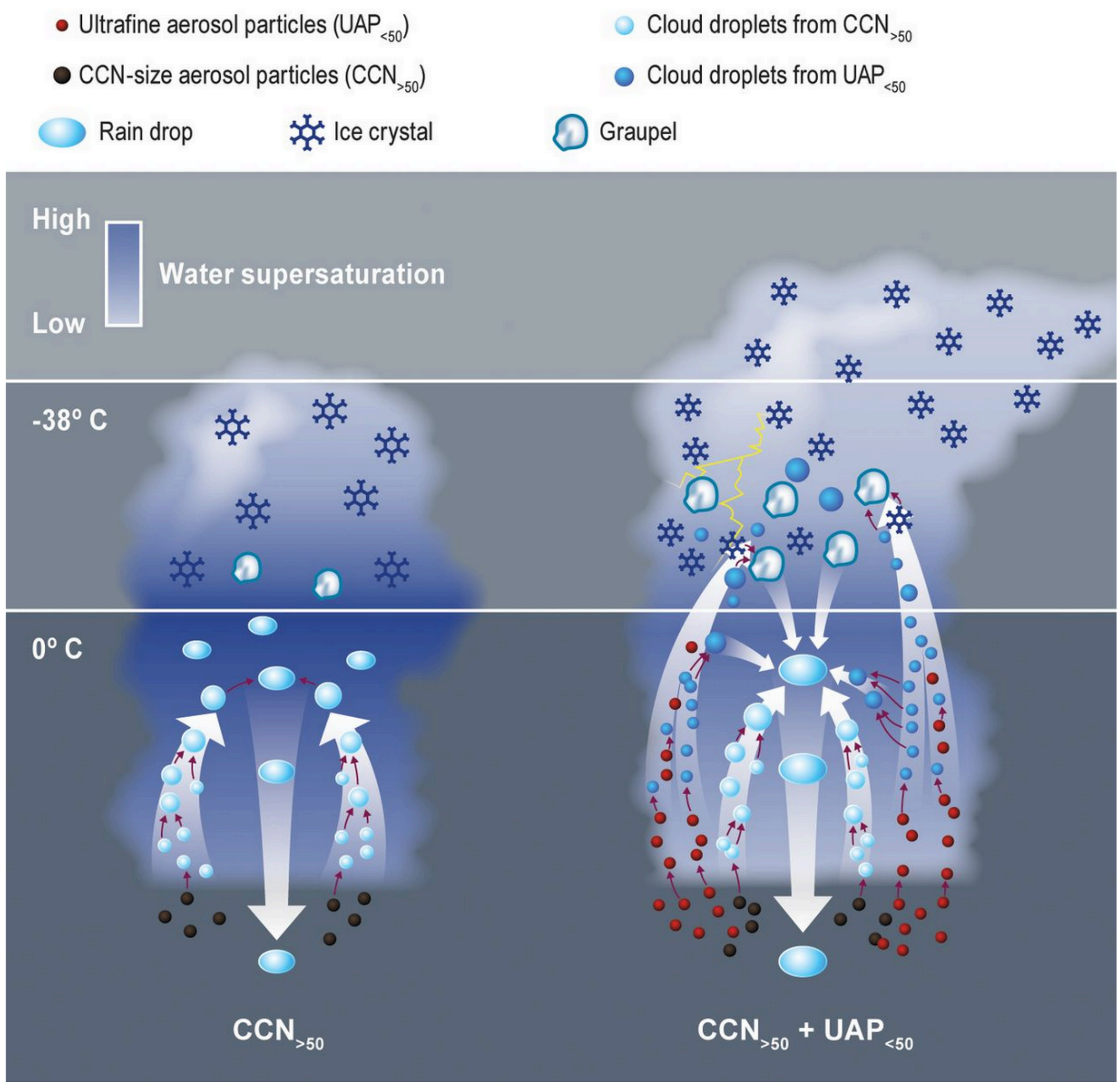

(Effect of UAP on cloud and precipitation formation; from 15)

This effect contributes to the creation of precipitation in nature, however the exact physics of cloud formation is incredibly difficult to replicate due to the multitude of effects that need to be modeled on relatively large scales for a huge amount of very tiny particles. Thus while we cannot quantify this and many other effects in any precise way we will attempt to take all effects into account as best we can.

The process described above is only one example of how precipitation is driven in nature. Convection can increase from any large temperature difference from the mean, see examples such as(16). However cooling, including from aerosols as we will see shortly, contributes to precipitation far more often than warming; which has a tendency towards stabilizing atmospheric movement and increasing local humidity in the average scenario, rather than decreasing it into clouds and then precipitation.

(a)To show that rapid local negative radiative forcing can lead to a short term increase in precipitation we can look to various models. One model studying radiative forcing effects of ozone, negative, finds that the short term cooling response shows an increase in precipitation globally (17). A review of radiative cooling to hydrological response shows the same thing, though in this case the weak negative response shows a very weak positive correlation with precipitation. That being said, the overall curve is an exponential one trending very clearly toward lower temperature being positively correlated with higher precipitation in the short term(18). This makes sense as we'll see just below.

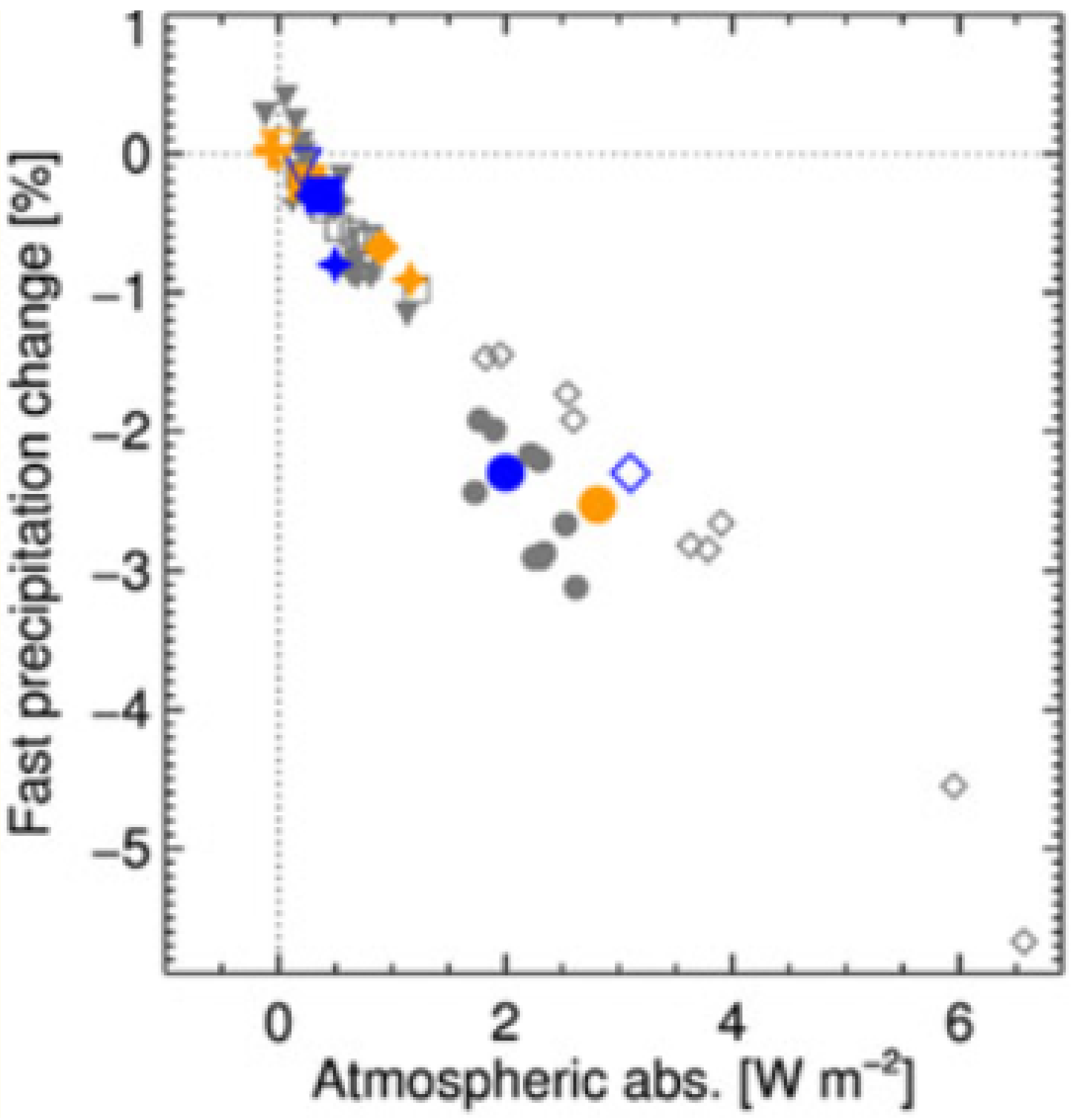

(From 18) *We can visualize a clear trend, both inversely correlating fast precipitation change and atmospheric absorption of various aerosols, and that the trend tends towards an exponential function as it goes into negative absorption/positive precipitation.*

Why precipitation tends towards a non linear response is elementary. The previously mentioned "DRH", the temperature at which water comes out of solution locally, must be reached before clouds may form. Only after this point can potential precipitation start to accumulate, while raising temperature above that will inhibit precipitation formation exponentially, and in reverse more water forced out of solution corresponds to more precipitation potentially seen on the ground. This is confirmed for us as we'll see below, where the nonlinear curve continues into cooling / positive precipitation response.

Thus we can show fundamentally how clouds form and ultimately lead toward precipitation; and that this can come from cooling and rapid negative radiative forcing. This task is what our aerosols will aim to accomplish. However before we define the cooling properties of our aerosols we will be concentrating on their safety. The two are linked, our aerosol must generate cooling (in the average scenario) for us to see precipitation, and our aerosol must be environmentally friendly in order to be considered for deployment. We choose to focus on safety first as that is often a paramount concern seen in atmospheric alteration projects.

## Safety Concerns

Our primary aerosol is calcium carbonate, a compound without any safety concerns as already discussed, however we also need to note that chemistry interactions are not the major concern here, as the mean particle lifetime of a tropospheric aerosol globally is just 3-7 days (19). Most aerosols in the troposphere, our area of deployment, are rained out fairly quickly.

But can we be sure of Calcium carbonate? Even when studied specifically for inhalation on humans no effect of concern has been found (20). With the composition of the earth's crust being around 4% calcium carbonate(21) of one form or another all recent life on earth should have evolved to be in regular contact with the substance to some degree; a logical leap but a solid one. Given that a large fraction of mineral dust is $CaCO_3$(22) with empirical observations to back up this up(23) and studies finding no environmental hazard from the substance(24) beyond making vegetation temporarily unpalatable to deer when used in vastly higher densities than we intend to deploy, we find it safe to assume that the proposed small scale temporary deployments of $CaCO_3$ in similar densities to those already encountered regularly around the world will be safe.

Thus while we find zero concern for safety of aerosolizing calcium carbonate artificially, as it is already aerosolized the world over regularly and we are simply adding relatively low amounts per square meter more over a given area and time, it is not our only compound. Low hygroscopic growth factor of calcium carbonate brings into question how successful a deployment will be before it is tested. Therefore as a backup we also propose "Sea Salts".

"Sea Salts" are the result of high wind speeds and crashing waves spraying salts, most relevant to us sodium chloride (NaCl) into the air as an aerosol, these are carried by winds far inland with extreme regularity. For our purposes NaCl has both a higher albedo than $CaCO_3$ and a very large hygroscopic growth factor, large enough that it would in effect double as both a cloud seeding compound and a cooling compound. However we need to demonstrate that this is safe to use as an aerosol, even if it is a naturally occurring phenomena. The primary concern stems from atmospheric deposition, which is to say aerosols blowing onto surfaces or falling as solute in rain.

As established above, and below, most coastal areas of the world experience large depositions of sea salt(19). Thus as we propose mimicking a naturally occurring phenomena whose deposition is safe in coastal areas. In doing so we should alleviate major health and environmental concerns in regards to those specific places; which suggests rapid practical testing in coastal areas should be accomplishable. Inland freshwater regions unadapted to such conditions are a different concern. The resulting potential damage therein is specific to the compounds used and areas it’s used in, which we’ll be discussing.

Aerosol reactions will be briefly mentioned here. Reactions from marine salt compounds have been studied in regards to long term radiative forcing scenarios, which found that resulting products can have an impact on the ozone layer among other concerns(19). However as our proposal centers on small scale, low altitude deployment wherein the compounds are expected to be “scavenged” out in hydrate form via clouds and rain, discussed further down, we see these concerns as inapplicable to our proposal in any meaningful sense.

Next we’ll be giving a brief overview of sea salts and their general safety profile. The primary component is Sodium chloride (NaCl); which is the most well known and abundant sea salt. The basis of table salt, it is universal in human diets in some fashion. However toxicity concerns exist, while hypernatremia (salt poisoning) and salinization of soil are commonly known problems, multiple studies have found NaCl concentrations to be among the safest of the so called “sea salts”. One overview of available literature shows no significant damage to any tested fauna or flora below 250mg/L(20,21,22). As we will see this is an extremely high, and unlikely, concentration for our proposal to reach even in the worst case. See further down for more discussion.

Other sea salts were considered, but were found to have less desirable safety profiles. Magnesium Sulfate (MgSO4) is highly available, though it has a rather low median IC50 (inhibition concentration) of 4mg/L(23); it has a fairly safe profile for human contact, having been studied for makeup application(24) and used for medical treatment wherein only exposure over a gram/h intravenous has been recorded as even possibly toxic for adults in the short term(25). Still, the IC50 precludes it from use here, as well sulfates in general have aging products that can severely affect the ozone layer far beyond NaCl, exacerbating climate change among many other undesirable effects(26). Even in short term deployment these may be undesirable.

In fact most salts are found to have lower PNEC (predicted no effect concentration) below sodium chloride, which has a PNEC of .5mmol/L (27). Thus for our purposes we’ll be sticking with Sodium Chloride as our primary aerosol. However the PNEC level in some strictly freshwater organisms is concerning. This should reinforce that deployment of our proposal should only take place over regions that already experience high chloride deposition (most chloride comes from NaCl) and those regions are most often coastal.

Concerning the area of deployment, no single relationship between coastal distance and atmospheric deposition can be defined. Thus regional deployment should be examined on a case by case scenario. However we find in studies of areas such as Sweden a single storm incident can deposit over half a gram of chloride over 50km from the coast, well within our proposed operational parameters(28).

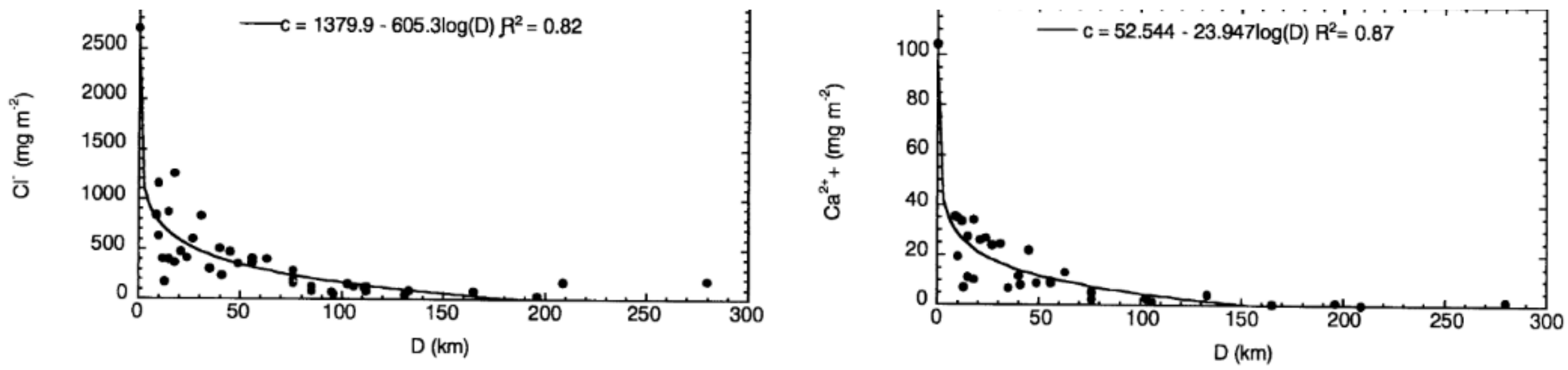

Fig. 6. Ion deposition to Pine needles across southern Sweden during a gale occasion in February 1989.

(from 28)

The exponential loss of chloride deposition as we move away from the coast confirms that chloride aerosols don't last long, or travel too far en masse, from their initial starting point. However one storm is not enough, thus we can find that deposition of cation for the entirety of Britain confirms that regular deposition of sea salts far inland occurs over the entirety of a large land mass depending on the prevailing winds(29). Thus we find it probable that sea salts could be used safely in many coastal deployment areas, however these would have to be decided on a case by case basis dependent on the already present averages of NaCl.

Finally we are limited to neither of these aerosols necessarily, but needed to pick two to concentrate on for the sake of publishing. Other compounds such as Kaolinite($Al_2Si_2O_5(OH)_4$) (30) are just as available cheaply, are present in every day dust around the world and thus assumed not to be harmful, and have high albedo to achieve effective radiative forcing; however as we are not currently writing a survey of all possible compounds that could affect local radiative forcing we'll concentrate on our two established ones for the paper.

Cooling Potential and CCN

(b)Calcium Carbonate(CaCO3) is highly available, and has an essentially non-existent toxicological profile below concentrations not relevant to this proposal. However CaCO3 has a severely low hygroscopic growth factor(31). While DHR (diurnal humidity range) activity remains low, impact scavenging of CaCO3 could cause already formed droplets to plateau in size instead of experiencing continued growth. Empirical studies of CaCO3 laden dust storms have found high correlation with lowered precipitation, even absent higher DHR aging products(32). Thus we find it likely that CaCO3 could inhibit precipitation if deployed in an area that mixed with rising air that creates the precipitation we are seeking. Thus deployment of CaCO3 would need to target a low enough altitude to ensure it does not last long enough in the atmosphere to cause a noticeable impact, but high enough that it does not significantly affect induced precipitation.

Fortunately we also note our proposed methods can be used to cool a given area without concern for precipitation in the case of catastrophic heat waves. As shown, in the low atmosphere calcium carbonate produces cooling without much fear of ozone loss(33). While other papers have questioned these reactions higher in the atmosphere, as we are deploying only in the troposphere where the compound will be removed via precipitation in a matter of weeks we see that there's acceptably minimal impact on the environment from low tropospheric deployment. Careful schemes of deployment may show that calcium carbonate can be used for creation of precipitation, but this will be up to future research.

(c)For creation of precipitation, most likely rain though there's no reason snow could not be affected as well, CaCO3 would likely need to be targeted so to be "above" a given cloud formation, cooling it to the point of precipitation formation while avoiding low hygroscopic growth factor limitations. As cloud microphysics is a computationally fantastically intense process on the scales proposed practical tests may be the best way to assess how plausible this is.

Thus our proposal remains firmly with calcium carbonate as a hopeful first compound whose limitations in precipitation formation can hopefully be overcome, but sodium chloride as a backup or complementary option is being kept in mind, a common compound with well studied safety profiles and an effective cooling capacity. With safety established we turn to cooling.

So how are these compounds to be aerosolized and used for cooling? Similar to stratospheric aerosol injection and marine cloud brightening we propose to use modified commercial aircraft to spray the compounds in an aerosol over a target area. Our method is to use negative radiative forcing, as in (3), however we'll be changing the goal from long term cooling to short term cooling to induce precipitation. Our next steps are to characterize when precipitation starts to form as a function of temperature/humidity, then the expected cooling versus of our compounds.

(d)For DHR (diurnal humidity range), the relative humidity at which we can expect pure NaCl to start gathering water from the atmosphere, we find 75.3% of relative humidity is needed(12). This is high compared to other available aerosols as we'll see. With a high DRH other aerosols will have started cloud formation before the relevant compound, and thus droplet size should be rather large by the time NaCl begins its own process. Thus NaCl should not, in most scenarios, starve available aerosols of usable water, as seen in previously mentioned scenarios, see below for confirmation. Further, should it be necessary, we see very close relatives of NaCl showing a much lower DRH, thus suggesting cheaply modified NaCl can be used if such a property is desired. Below is a chart of both pure and common relatives of NaCl in terms of DRH.

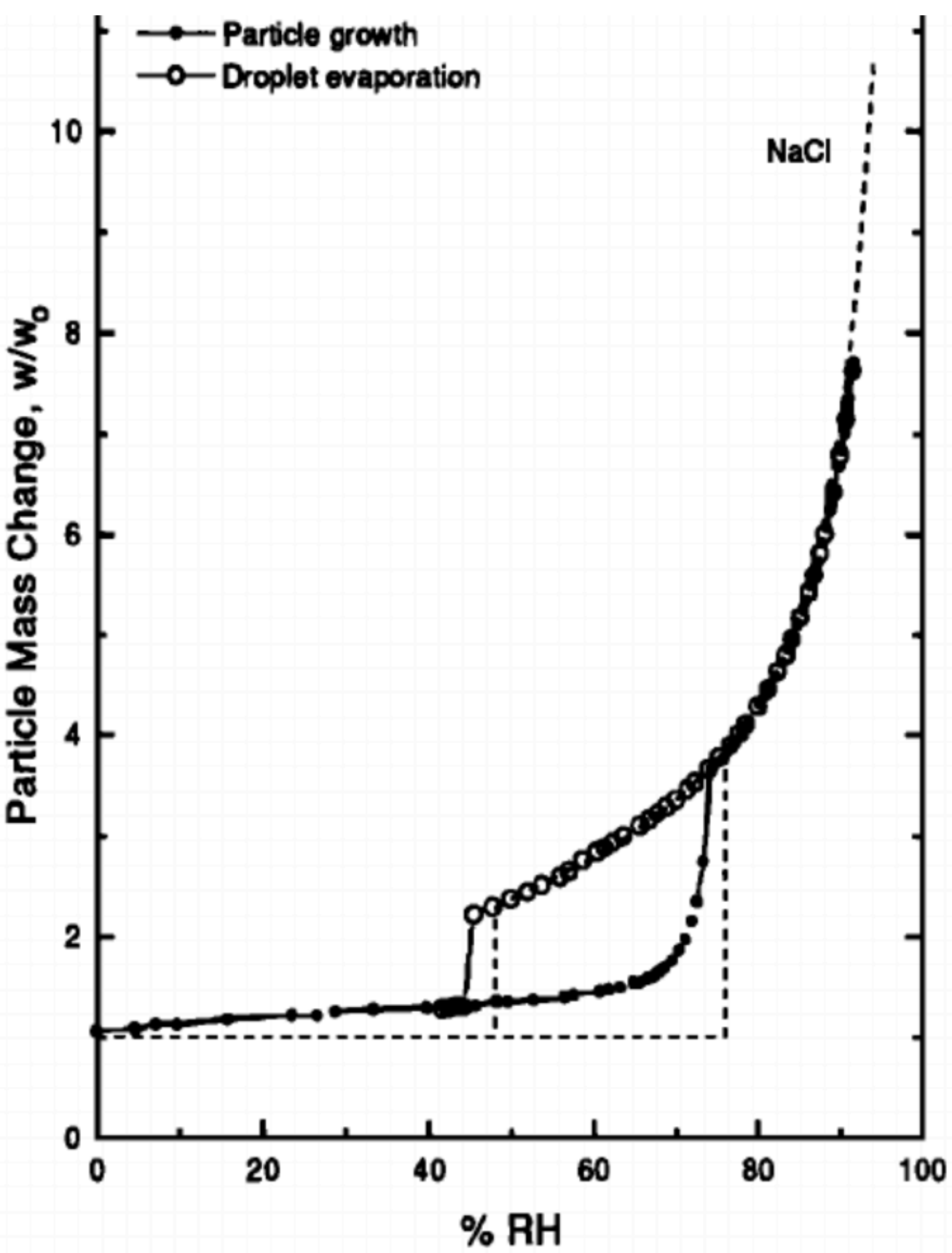

**Figure 1.** Phase transformation, growth, and evaporation of a sea salt particle as a function of relative humidity. For comparison, the hydration behavior of a pure NaCl particle is illustrated as dashed curves and lines.

(From 35)

We can also see growth factor has no limit for pure NaCl, another excellent property for our use case. Please note that particle sizes < 150nm can cause differences in hygroscopic growth, with an increase with decreased sized noted(36). However we don't plan to use particles this small.

(e)For our CaCO3 scenario we find that relative humidity/temperature are more relevant coefficients than exact aerosols in a typical scenario(37). Thus in both this scenario and sea salts we need to concentrate first on cooling, and second on aerosol mix, assuming a typical aerosol mixture and there are no particular suppression effects present. This does mean we need water to already be present in the atmosphere of course, but that was a given.

Radiative forcing is next, this is the amount of radiation, in units $wm^{-2}$ scattered back into outer space, or rather cooling capacity. This calculation is anything but straightforward. It is often done with "clear sky" and "total sky" calculations, referring to a clear day and a cloudy day. Total radiative forcing is lower once clouds start reflecting sunlight themselves, leaving less radiation for the aerosol to scatter. Further, radiative forcing is affected by "sun zenith", which is to say the higher the sun is in the sky the more radiation will be scattered directly into space, whereas the lower it is the more will be reflected towards the ground. Thus we can expect latitude, time of day, and season to all affect our expected cooling capacity visa vis radiative forcing.

For CaCO3 (calcite) we find a range of different expectations, however direct observations of dust storms often composed chiefly of calcite have found direct satellite observations of dust storms causing as much as -91 - 114*w/m^2* forcings, far above what we'd expect to target in any given scenario(38). The aerosol burdens for negative forcings appear to be on the orders of mg per $m^{-2}$(30) , so we'll take the rough estimate of 4mg -1w/m^2, or rather per gram -250$w^{-1}/g^{-1}(m^{-2})$. Though it's hard to tell from direct observations as the aerosols are mixed fractions, theoretical calculation back up that high calcite fraction is one of the most negative available compounds found in common dust aerosols(34). A more precise value for this will have to wait for actual tests.

As for NaCl a normalized forcing of -88$w^{-1}/g^{-1}(m^{-2})$(39) for "clear sky" which is our expected starting scenario. Further it notes the optimal particle size for reflecting visible wavelengths is around 500nm. One caveat is that "total sky" or cloudy scenarios see this forcing cut in half, though this does mean even once clouds start forming cooling should continue even above normal cloud scenarios.

One concern, as noted in (40) is that radiative forcing must be relative to already existing albedo. The most notable finding therein being that deserts can have a higher albedo than NaCl, meaning radiative forcing can become positive relative to average for sea salt aerosols over regions as such. With snow having a very high albedo we can also assume much the same will be the case there. The general case we will assume however is that surrounding albedo is the earth average of around 0.3, and thus overall forcing will be negative. This does reinforce the need to evaluate each potential deployment area separately.

Another note is that NaCl has differing properties due to its shape. This can include gathering small amounts of water out of solution at an RH of around 46% such as seen in the graph above, and changes in refractive index, and thus ultimately radiative forcing, relative to particle size(41). Our previous estimate for forcing was validated against empirical data within a decent error bar, and thus stands as a reference for this paper. However sourcing of both CaCO3 and NaCl should affect both DRH water adsorption and radiative forcing dependent on shape. Whether this can be used to favorably affect the CaCO3 water droplet plateau will be left to future work.

Finally, we note that bare NaCl has a high reflectance spectra across solar radiation wavelengths, including infrared and optical(42). This is excellent for cooling purposes. However as soon as water absorption starts, which has been shown to start as low as an RH of 46% dependent on the morphology of NaCL used, this albedo can be altered as water can absorb infrared more readily. This should change radiative forcing, potentially before major cloud formation. However, the overall impact is still shown to be negative total forcing and a cooling effect(43). This is noted here to emphasize the complexity of attempting to model the proposed process in detail, which will affect how we model and what models we choose, as we will see below.

Next we'll convert the cooling capacity from watts to temperature. Switching around our estimate from above we find $1w/m^{-2}$ RF needs .0113g for NaCL, whereas we approximate 0.004g per $1w/m^{-2}$ for CaCO3 in "clear sky" conditions. Herein we'll describe the conversion with the heat transfer coefficient

$$\alpha W m^{-2} = \Delta K^{-1}$$

This directly relates watt per meter squared $\alpha$ to temperature difference in Kelvin. In a basic model $\alpha = 0.5$ as predicted by(44). However a different, more detailed set of modelling based on the aerosol injection technique "marine cloud brightening" includes cloud radiative forcing and other effects, and found $\alpha <= 1$ dependent on specific conditions(45).

Other estimates swing wildly as well, one studying sea salt specifically in a global case with climate change scenarios included found $\alpha = 1.44$ (46). The end result is that $\alpha$ seems to have no easy fixed point, and should vary based climatic conditions, wavelength dependent response, and other coefficients. Still, we can use $\alpha = 1$ in our calculations below for simplicity's sake as it is within our estimate range. Next we'll discuss the general altitude we may wish to deploy at, and other non radiative concerns.

Briefly, An empirical study confirms the basic model of a decreasing water vapour content with increasing altitude(48). As most of the water vapour we want to condense into precipitation is in low altitudes this limits the need or desire for a cooling effect to these same altitudes. There are concerns over sea salt type reactive chlorine and it's potential O3 interactions(46). However as discussed above we expect our aerosols to have a short lived lifespan, and as we'll see below we plan to deploy relatively little aerosol versus our expected effect.

As for the radiative cooling potential of calcium carbonate, a useful paper has already been presented showing cooling of several degree celsius is eminently reachable(47). Whether this can be adapted to create precipitation by say, increasing deployment height with the intention of creating an upwelling that produces sufficient droplet size before droplet size mitigation is a topic for future research.

**Results**

Finally we get to modeling both the results of our methods and initial estimates of financial outlay and end results of a theoretical deployment.

Our first task will be to validate our previous set of assumptions via empirical data. That is to say, that cooling will increase air pressure/lower dew point (mbar/C) until such time as humidity drops near or below the dew point; at which point cloud formation should begin(a). This will continue in a feedback loop until humidity is exhausted versus the local aerosol/DRH availability compared to the dew point; and hopefully, precipitation begins during or shortly after. We see exactly this here, validated against empirical data(49)

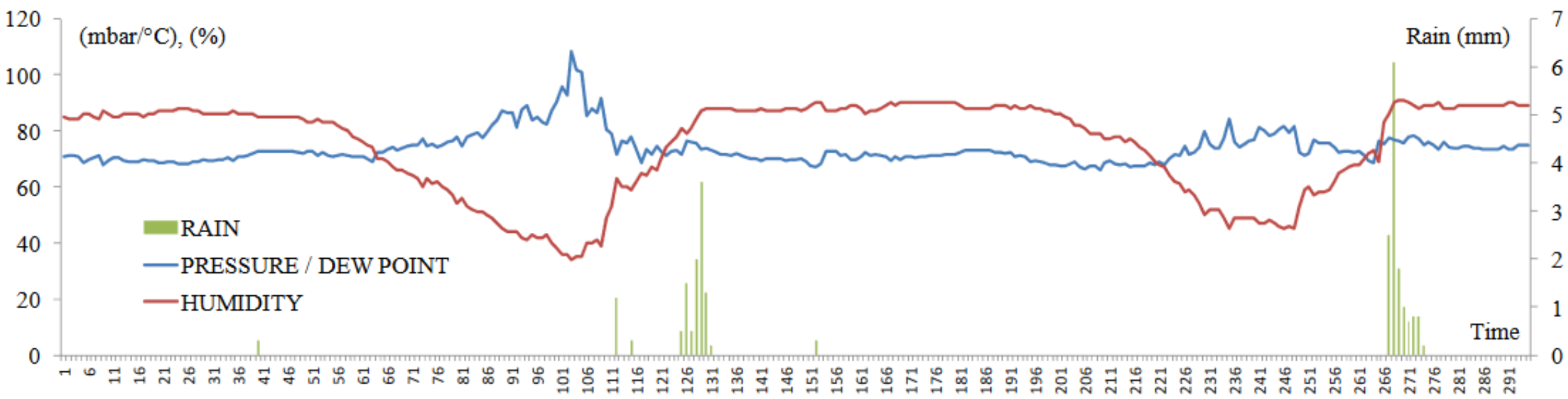

**Figure 10.** Results of applying the CRHUDA model to Milenio data on August 16, 2014 (max depth 12.1 mm in 1 h; max intensity 42.3 mm/h; 37-year return period).

(from 50)

This algorithm gives us a simple relationship between the correlated pressure/dew point and humidity. We can see the uplift reaction from up above, lifting moisture into the air, forming cloud as the dew point increases until it reaches saturation even as humidity decreases, then the resulting rain is formed and equalizes the two again. This plot also demonstrates our limitations, we need water to be present in the atmosphere before we can wring it out, a high relative humidity is observed before precipitation forms. We'll address this further down below.

Next we'll address CCN, a generally much more complicated topic. For CaCO3 we already found that this can suppress cloud formation from direct observation as we discussed back in (b) and (c). Thus we'll address that further down below. For the moment we can discuss sea salts (d) further. For this we'll turn to a CCN Parcel simulation(51).

Our aerosol coefficients for the simulation are hygroscopicity(often denoted with κ), a reference can be found from(52) where the average k (kappa) is given as 0.42 for one set of saturation conditions in Taiwan, which we'll be using. It is worth noting that this parameter has a very complex link to other parameters ranging from particle size, exact dimensions, and etc. Other parameters are mean size, std. deviation, and mean concentration, all things we would be choosing in a deployment for our aerosol but not the naturally present aerosols.

Our starting parameters are found in fig 1 and 2 for before and after our theoretical aerosol injection respectively, we show a pure sodium chloride with enough injected (over a small vertical theoretical column, but as we'll see this is arbitrary) to affect an extreme 3 degree kelvin delta, and the resulting parcel simulations in fig 3 and 4 show our resulting droplet size. We'll discuss how droplet size can be used to predict precipitation after.

```
    output_dt: 1.0
    solver_dt: 10.0
    t_end: 9999.0
    terminate: true
    terminate_depth: 10.0

: Initialize the model with two aerosol species:
.nitial_aerosol:

    # 1) a mixed mode of small particles, with 50 bins
    - name: mos
      distribution: lognormal
      distribution_args: { mu: 0.073, N: 1202, sigma: 1.05 }
      kappa: 0.42
      bins: 250

: Set the model initial conditions
.nitial_conditions:
    temperature: 285.15 # K
    relative_humidity: 0.65 # %, as a fraction
    pressure: 85000.0 # Pa
    updraft_speed: 0.44 # m/s
```

**Figure 1**

```
# Initialize the model with two aerosol species:
initial_aerosol:

    # 1) a mixed mode of small particles, with 50 bins
    - name: mos
      distribution: lognormal
      distribution_args: { mu: 0.073, N: 1202, sigma: 1.05 }
      kappa: 0.42
      bins: 250

 # 2) a sea salt mode, with 250 bins, and
    - name: sea salt
      distribution: lognormal
      distribution_args: { mu: 0.5, N: 800, sigma: 1.01 }
      kappa: 1.2
      bins: 250

# Set the model initial conditions
initial_conditions:
    temperature: 282.15 # K
    relative_humidity: 0.79 # %, as a fraction
    pressure: 85000.0 # Pa
    updraft_speed: 0.44 # m/s
```

**Figure 2**

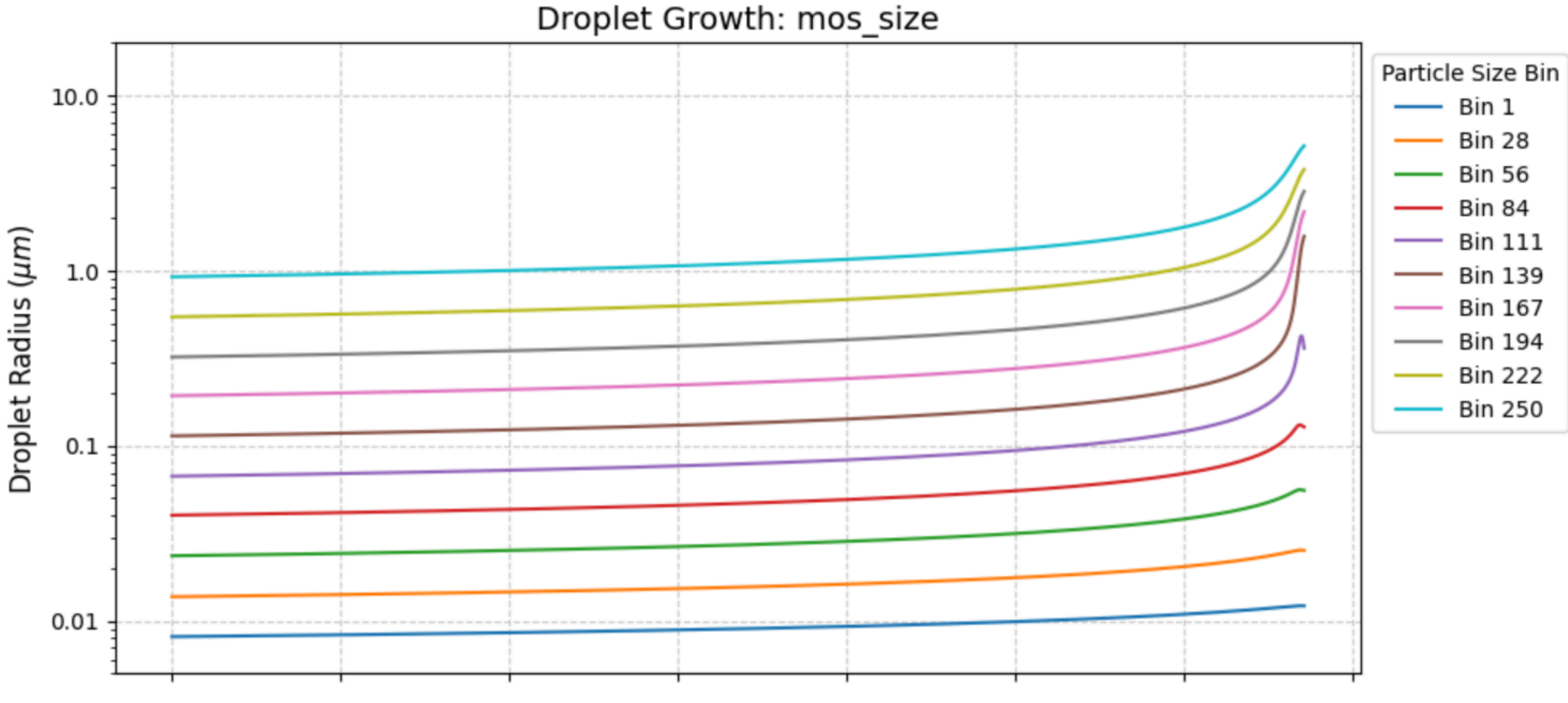

**Figure 3**

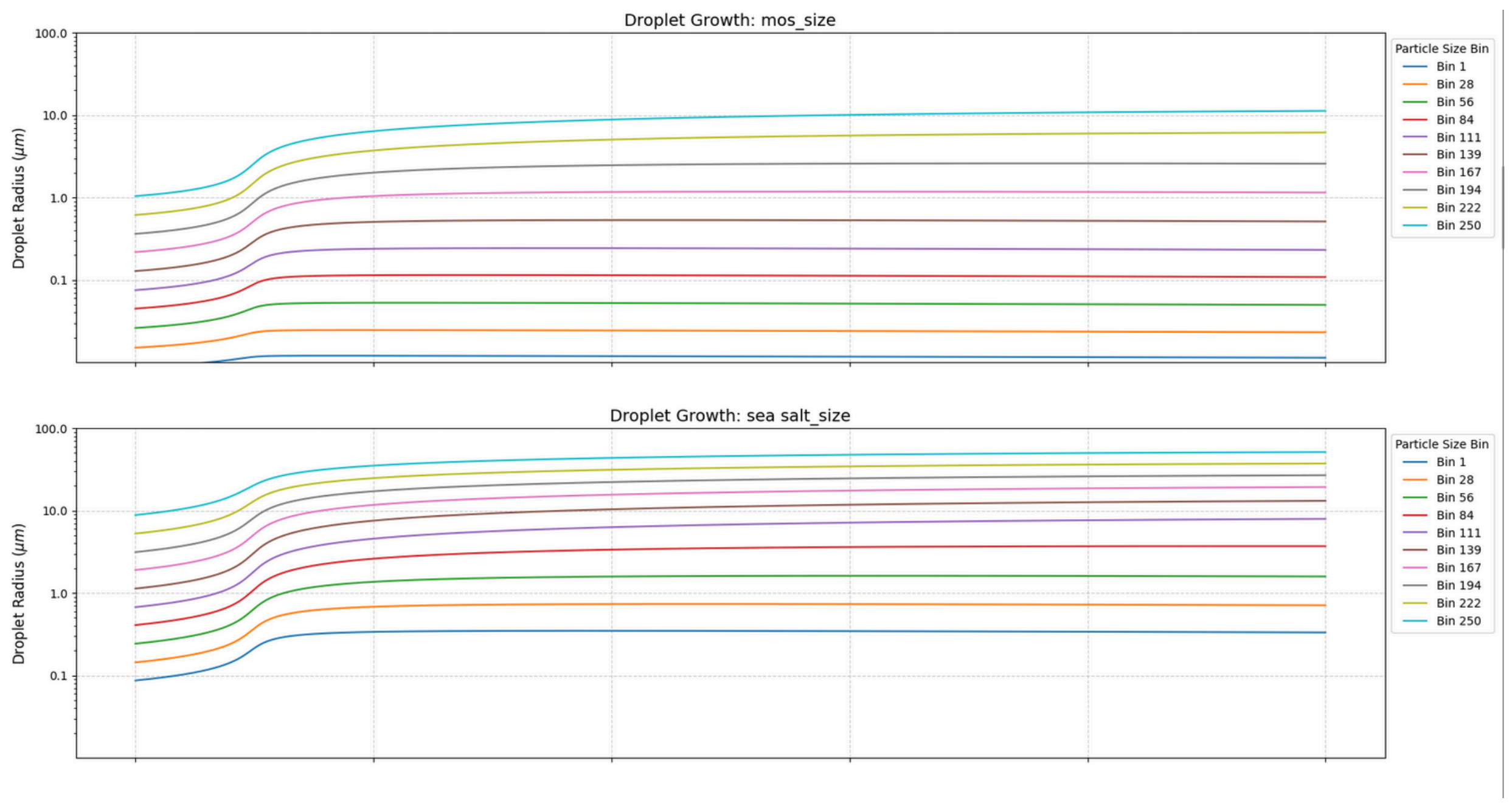

**Figure 4**

Time is on our X axis and droplet size in microns in on our Y. Our simulation assumes a rising parcel of cloud. Earlier we noted that CCN coming out of solution causes this very affect, and as we can see from our first scenario our droplet growth is around flat, which should not cause much of any exothermic reaction, whereas we start out with a clear slope upwards with our injection scenario (fig 4) that shows we can assume a rising parcel.

The other key results the sim we are looking for is droplet size, as noted in(53) we need not get above 30 micron droplet size before start seeing precipitation on average. However what this parcel simulation does not account for is droplet convergence, only hygroscopic growth.

However looking at (54) we can find a time series wherein modelling droplet collision shows that at t 10 minutes the start of droplet radius around or above 10 microns, as our injection scenario shows, will quickly coalesce into droplets far above our 30 micron cutoff just 10 minutes later

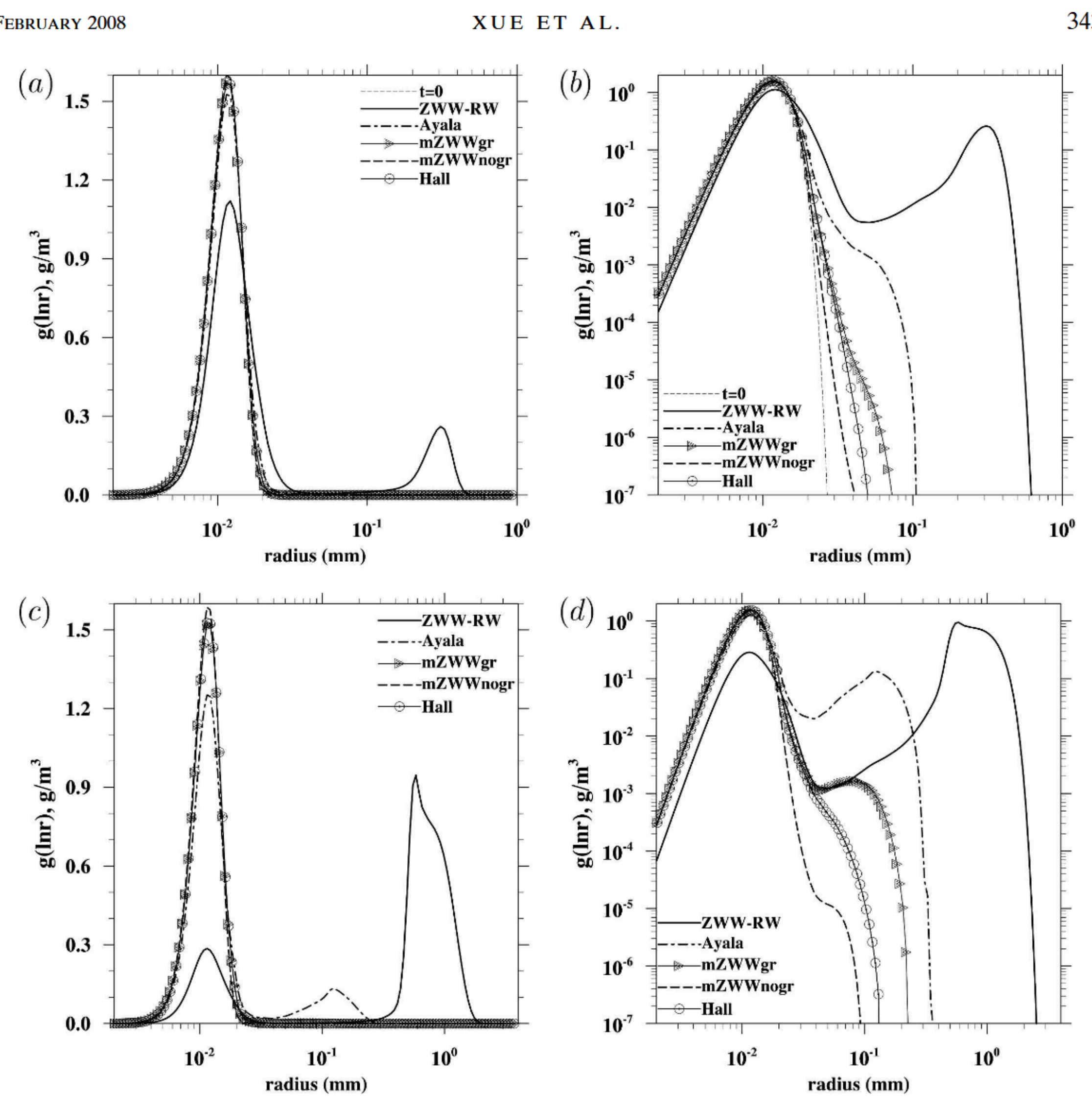

FIG. 10. Mass density distributions at (a) $t$ = 10 min, (b) $t$ = 10 min on logarithmic scale, (c) $t$ = 20 min, and (d) $t$ = 20 min on logarithmic scale. The flow dissipation rate and rms velocity fluctuation are $\varepsilon$ = 300 $cm^2 s^{-3}$ and $u'$ = 202 cm $s^{-1}$.

from (54)

Thus we can show that combining a simulation of hygroscopic growth with real world empirical data and a well founded assumption from previous research that we can achieve exothermic cloud lift into a droplet growth regime that, going by previous research, should achieve droplet growth sufficient to cause precipitation.

Our overall results lead us to expect precipitation, given the expected parameter shifts of CCN and temperature, over a simplified model. But how certain can we be of our results? Here we turn to (55) and see a different simulations showing enhanced precipitation creation in marine weather systems from sea salt aerosols, done by an entirely independent research team. This supports our theory of using such to induce precipitation.

With positive results from our cloud simulation concerning sea salts how might we address CaCO3 use? As already discussed back in (e) we show that temperature is a bigger factor than specific CCN for cloud condensation here. So we need to estimate how much we might need to change the temperature in a real world scenario. At the same time we can confirm, in yet another simulation regime, that our basic premise of lowering the temperature of a region producing precipitation is sound.

Let's use an ML algorithm trained on historical Australian weather(56). The drawbacks of ML as such are that we are more or less making an educated guess, the benefits are that the guess has implied education about real world CCN in the given area and we need only input temperature changes, all other effects linear or nonlinear should be decently approximated.

We'll input the data for a scenario where the need for rain is particularly high: June of 2019 on the Australian Southeast Coast saw devastating fires. We'll start on April 1st in Sydney, saying we need rain then and there as the unusually dry season is in progress. Here's the weather data from that day:

## Summary

| **Temperature (°F)** | Actual | Historic Avg. | Record |
| --- | --- | --- | --- |
| High Temp | 68 | 75.5 | -- |
| Low Temp | 55 | 60.1 | -- |
| Day Average Temp | 63.52 | -- | - |
| **Precipitation (in)** | Actual | Historic Avg. | Record |
| Precipitation (past 24 hours from 14:00:00) | 0.00 | 0.20 | - |
| **Dew Point (°F)** | Actual | Historic Avg. | Record |
| Dew Point | 53.89 | - | - |
| High | 61 | - | - |
| Low | 45 | - | - |
| Average | 53.89 | - | - |
| **Wind (mph)** | Actual | Historic Avg. | Record |
| Max Wind Speed | 21 | - | - |
| Visibility | 6 | - | - |
| **Sea Level Pressure (in)** | Actual | Historic Avg. | Record |
| Sea Level Pressure | 30.28 | - | - |

(Acquired from:57)

If we use our ML algorithm to predict the weather tomorrow, accounting for the lowering temperature and relative humidity from the starting point, we find we need to lower the temperature by just 2.5 degrees celcius to change our prediction to rain, all other things being equal. This is well within our estimated grasp and goes along with our observed rainfall above, but let's define what we might reasonably expect to accomplish. Thus we can firmly show, from previous research, and from two different simulations ourselves, that we can anticipate success from our method.

So how might we go about, reasonably, spreading our aerosol over a given area? An Airtractor 802, specifically designed for spreading aerosols, has a dedicated hopper of 3k liters(58) (dependent on model). Calcite has a density of 2.7g/cm^3, but would obviously be a lot of air when ground up into a fine dust, this matters as our useful load (aerosol mostly, in this case) is 4 tonnes. If it was solid Calcite we'd be weight limited, with over 6 tonnes of calcite. So let's limit ourselves to a weight of 4 tonnes of Calcite.

Going on to our normalized forcings, even if we cut our estimations drastically, to -100w/g per meter, we find 1 tonnes would deliver -100 watts cooling for a square kilometer. At our target of roughly -2.5w we find we could cover 160 kilometers squared with one plane load of a rather small plane. Just under 13 by 13 kilometers sounds like an entirely reasonable target for a purchasable single engine plane. A quick search of Alibaba, which we won't bother linking to, suggests industry grade calcite powder is currently $350 USD per tonne, an entirely reasonable cost for delivering rain. Even if we half, or quarter, again our total effective assumption of aerosol weight to degrees celsius to rain, another one to four plane loads breaches neither our safety margins nor budgetary margins. By all estimations we can produce rain at reasonable cost.

But this is under specific circumstances. We have limitations, to demonstrate we turn to: Santa Rosa, Ca on August 16th, 2025. In the morning humidity is high, dew point is close, and we might need only a few degrees celsius to start our process of water coming out of the atmosphere:

## Daily Observations

| Time | Temperature | Dew Point | Humidity | Wind | Wind Speed | Wind Gust | Pressure | Precip. | Condition |
|---|---|---|---|---|---|---|---|---|---|
| 12:53 AM | 62 °F | 57 °F | 84 % | SSE | 5 mph | 0 mph | 29.75 in | 0.0 in | Fair |
| 1:53 AM | 61 °F | 57 °F | 87 % | SE | 5 mph | 0 mph | 29.74 in | 0.0 in | Fair |
| 2:53 AM | 60 °F | 57 °F | 90 % | SSE | 3 mph | 0 mph | 29.74 in | 0.0 in | Fair |
| 3:53 AM | 59 °F | 57 °F | 93 % | SSE | 5 mph | 0 mph | 29.75 in | 0.0 in | Fair |
| 4:53 AM | 59 °F | 57 °F | 93 % | SSE | 3 mph | 0 mph | 29.75 in | 0.0 in | Fair |
| 5:53 AM | 58 °F | 56 °F | 93 % | S | 3 mph | 0 mph | 29.76 in | 0.0 in | Fair |
| 6:53 AM | 58 °F | 56 °F | 93 % | SSE | 5 mph | 0 mph | 29.77 in | 0.0 in | Fair |
| 7:53 AM | 62 °F | 58 °F | 86 % | S | 3 mph | 0 mph | 29.79 in | 0.0 in | Fair |
| 8:53 AM | 67 °F | 59 °F | 76 % | SE | 5 mph | 0 mph | 29.80 in | 0.0 in | Fair |
| 9:53 AM | 72 °F | 59 °F | 64 % | SSE | 7 mph | 0 mph | 29.80 in | 0.0 in | Fair |
| 10:53 AM | 77 °F | 60 °F | 56 % | VAR | 7 mph | 0 mph | 29.80 in | 0.0 in | Fair |
| 11:53 AM | 82 °F | 61 °F | 49 % | SSE | 7 mph | 0 mph | 29.80 in | 0.0 in | Fair |
| 12:53 PM | 85 °F | 61 °F | 44 % | SSE | 8 mph | 0 mph | 29.79 in | 0.0 in | Fair |
| 1:53 PM | 86 °F | 62 °F | 44 % | S | 10 mph | 17 mph | 29.78 in | 0.0 in | Fair |

(from weather underground, see above)

However in the same place just a few days later we can see that by 10am we are at just in the 60's in terms of relative humidity, meaning getting rainfall out of this day is much harder.

## Daily Observations

| Time | Temperature | Dew Point | Humidity | Wind | Wind Speed | Wind Gust | Pressure | Precip. | Condition |
|---|---|---|---|---|---|---|---|---|---|
| 12:53 AM | 53 °F | 50 °F | 89 % | SSW | 3 mph | 0 mph | 29.83 in | 0.0 in | Fair |
| 1:53 AM | 53 °F | 50 °F | 89 % | CALM | 0 mph | 0 mph | 29.83 in | 0.0 in | Fair |
| 2:53 AM | 51 °F | 49 °F | 92 % | CALM | 0 mph | 0 mph | 29.84 in | 0.0 in | Fair |
| 3:53 AM | 51 °F | 48 °F | 89 % | CALM | 0 mph | 0 mph | 29.84 in | 0.0 in | Fair |
| 4:53 AM | 50 °F | 48 °F | 93 % | CALM | 0 mph | 0 mph | 29.84 in | 0.0 in | Fair |
| 5:53 AM | 50 °F | 48 °F | 93 % | CALM | 0 mph | 0 mph | 29.86 in | 0.0 in | Fair |
| 6:53 AM | 47 °F | 46 °F | 97 % | W | 3 mph | 0 mph | 29.88 in | 0.0 in | Fair |
| 7:53 AM | 53 °F | 50 °F | 89 % | CALM | 0 mph | 0 mph | 29.89 in | 0.0 in | Fair |
| 8:53 AM | 59 °F | 50 °F | 72 % | CALM | 0 mph | 0 mph | 29.89 in | 0.0 in | Fair |
| 9:53 AM | 65 °F | 51 °F | 61 % | CALM | 0 mph | 0 mph | 29.89 in | 0.0 in | Fair |
| 10:53 AM | 71 °F | 50 °F | 47 % | W | 5 mph | 0 mph | 29.88 in | 0.0 in | Fair |
| 11:53 AM | 75 °F | 50 °F | 41 % | S | 7 mph | 0 mph | 29.87 in | 0.0 in | Fair |
| 12:53 PM | 78 °F | 48 °F | 35 % | VAR | 3 mph | 0 mph | 29.86 in | 0.0 in | Fair |

This is an example of our technique being reasonable in expectation for a real world scenario, but reliant on opportunity, as humidity drops to the 40's by mid morning. We can reasonably expect success, but will need to be ready to go when a large amount of precipitable water is in the air column, unlike desalination we cannot expect to produce water regardless of conditions.

Next we can turn to practical benefits: Generating rainfall is desirable, it can prevent wildfires, relieve water pressure on local flora and fauna, provide water for local crops, and more. But a strong motivation is also to provide drinking water for nearby residents. So how much drinking water can rainfall provide? Here we turn to a model of a reservoir in Taiwan, which provides estimates of the amount of rainfall lost to secondary sources versus how much ends up in the reservoir over its catchment area(59). The resulting equation is nonlinear, with a large and area and condition dependent strong initial loss. However after a relatively short interval, less than half of the time series of the storm studied, the amount of water going from rainfall to reservoir becomes so close to 100% that it falls below their error bar.

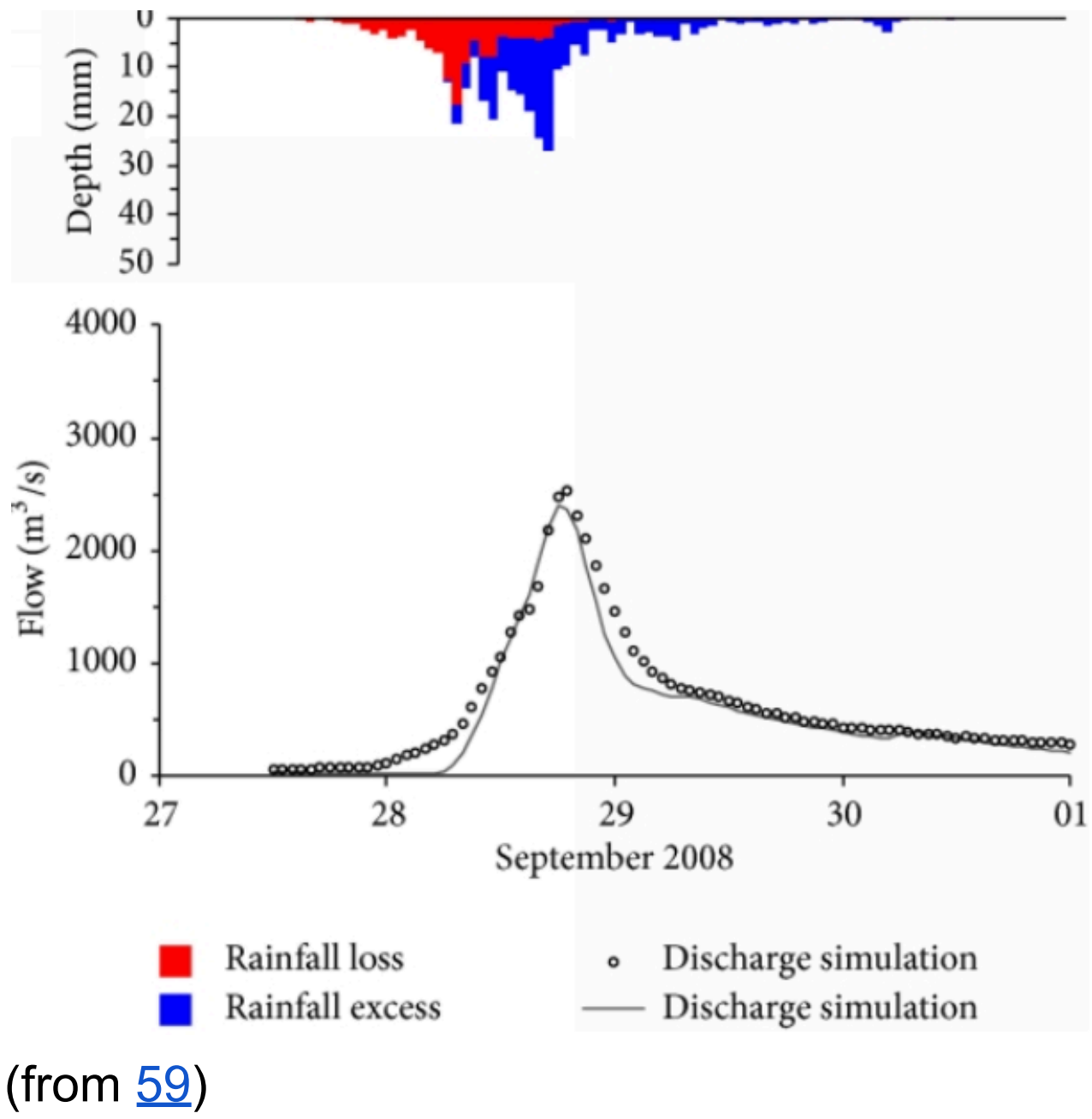

(from 59)

Other studied storms from the same reservoir estimate different loss curves, confirming that the equation is condition dependent even in the same reservoir catchment. However we can take some guidance from this without creating some more complicated model. Beyond around 30mm of precipitation over a given area the amount of water going into a reservoir sharply trends towards 100%, which will suit our rough cost estimate case study. Overall, the more rain we induce in a short timeframe (several days), the more cost efficient the resulting drinking water will be, this principle should hold regardless of area or conditions.

Of note, recent claims of successful precipitation inducement with cloud ionization(60) show a potential complementary technique(61) to the one proposed here. Cloud ionization is hypothesized to induce more droplet collisions arising from physical effects (see paper). As we propose to generate cloud drops and/or enhance droplet size, we see this effect as entirely complementary to our own proposal and both could be used in concert successfully if desired.

With our case study completed we feel confident in our conclusions, see below.

**Conclusion and Future Research**

We have shown that rapid cooling from appropriate aerosols could achieve artificial precipitation over a target area under two different simulation methods of our own and further research from other groups. Further we have shown that both CaCO3 and NaCl present promising candidates as an immediate, inexpensive, and under the right conditions safe aerosol for accomplishing this task. Considering the growing need for freshwater in many areas of the world, we see immediate further research and even practical testing as a target for this avenue of research.

In terms of future research there is much left to be done. Given that practical testing is unbeatable in terms of research value we suggest this should be considered. Beyond that there are many questions and optimizations left as well. The optimal particle size, choosing between CCN activity and radiative forcing, is yet to be determined. New aerosols that may safely be deployed, and those even with improved forcing and CCN properties is a highly promising avenue. We leave yet more avenues for future research up to the reader.